\documentclass[aps,onecolumn,superscriptaddress]{revtex4-2}
\usepackage{graphicx,color,setspace}
\usepackage[utf8]{inputenc}
\usepackage{mathtools}
\usepackage{tabularx}
\usepackage{enumerate}%
\usepackage{float}
\usepackage{ragged2e} %
\usepackage{subcaption}
\DeclareCaptionJustification{justified}{\justifying}

\usepackage[labelfont=bf]{caption}

\begin{document}

\author{\footnote{The author to whom correspondence should be addressed: marco.cappa@uniroma1.it} Marco Cappa}
\affiliation{Dipartimento di Fisica, Sapienza Universit\`{a} di Roma, P.le Aldo Moro 5, 00185 Rome, Italy}
\author{Francesco Sciortino}
\affiliation{Dipartimento di Fisica, Sapienza Universit\`{a} di Roma, P.le Aldo Moro 5, 00185 Rome, Italy}
\author{Lorenzo Rovigatti}
\affiliation{Dipartimento di Fisica, Sapienza Universit\`{a} di Roma, P.le Aldo Moro 5, 00185 Rome, Italy}

\title{A phase-field model for solutions of DNA-made particles}

\begin{abstract}
We present a phase-field model based on the Cahn-Hilliard equation to investigate the properties of phase separation in DNA nanostar systems. Leveraging a realistic free-energy functional derived from Wertheim theory, our model captures the thermodynamic properties of self-assembling DNA nanostars under various conditions. This approach allows for the study of both one-component and multi-component systems, including mixtures of different nanostar species and cross-linkers. Through numerical simulations, we demonstrate the model’s ability to replicate experimental observations, including liquid-liquid phase separation, surface tension variation, and the structural organisation of multi-component systems. Our results highlight the versatility and predictive power of the Cahn-Hilliard framework, particularly for complex systems where detailed simulations are computationally prohibitive. This work provides a robust foundation for studying DNA-based materials and their potential applications in nanotechnology and biophysics, including liquid-liquid phase separation in cellular environments.
\end{abstract}

\maketitle

\section{Introduction}
DNA constitutes a macromolecule of extraordinary interest from different points of view: its particular double helix conformation, combined with the hybridisation mechanism that allows the coupling between sequences of compatible nitrogenous bases~\cite{DNA}, makes it an instrument of choice for the synthesis of innovative and versatile materials in the field of materials science, and a powerful means of investigation and experimentation at the biophysical level~\cite{lorenzo}; in this sense, significant are the uses that have been made of it (even in aggregate form) to create nanomachines~\cite{nanomachines}, logical \textit{gates}~\cite{gates}, nanostructures~\cite{DNA_origami1, DNA_origami2,biobrick1, biobrick2}, to build vectors for \textit{drug-delivery}~\cite{drug_delivery} strategies, but also to investigate novel phenomena in soft matter systems~\cite{separation2,biffi,conrad2019increasing,conrad2022emulsion}. In particular, the sequence-specific pairing mechanisms of DNA makes it particularly suitable to study phase separation processes, as they allow for a careful control on the bonding properties of the system~\cite{lorenzo, biffi, conrad2022emulsion}. 

A class of DNA-based systems that has been used for this kind of investigation are specific DNA constructs known as nanostars, where the maximum number of bonds that each particle can create can be controlled by design. A nanostar of valence $\mathcal{V}$ is formed by $\mathcal{V}$ single DNA strands hybridised together and capable of binding to each other through sticky ends~\cite{lorenzo, biffi}; control on the system bonding modes is achieved through the selection of the base sequences making up the nanostar binding sites, so that only specific sticky ends can bind to each other.

Given the complexity of the DNA molecule, formed by long chains of nucleotides containing dozens of atoms, any atomistic descriptive model is unsuitable to characterise its thermodynamic properties: the high number of degrees of freedom would make the simulations extremely large and therefore unmanageable in terms of implementation and timing~\cite{mesoscale}. One possibility to simplify the problem consists in using particle-based coarse-grained models~\cite{coarse_grained}, which indeed makes it possible to access the timescales required to study collective phenomena~\cite{lorenzo2, lorenzo3}. However, phenomena such as phase separation and coarsening are still out of reach at such a level of description.

Here we tackle the problem by using a different approach based on the Cahn-Hilliard equation~\cite{ch}, which is a phase-field, partial differential equation describing the time evolution of the macroscopic order parameter density $\rho$, initially developed for the study of metal alloys, but in general suitable for the description of fluid mixtures with one or more components~\cite{ch, alloys}. Differently from previous attempts~\cite{diffusion}, based on schematic free-energy expressions, we describe the thermodynamics of the DNA nanostar systems in a realistic way making use of a mean-field free energy which has been developed to model self-assembling systems, and has been shown to qualitatively predict the properties of these DNA constructs~\cite{lorenzo}.

Our aim is to assess the validity and potential of this \textit{phase-field} model, as well as its predictive accuracy with respect to theoretical expectations and especially experimental results: the latter match is of particular interest because of the wide range of potential applications in biophysics involving nanostar-based systems, which include the possibility to mimic the liquid-liquid phase separation processes that are important for many cellular processes~\cite{condensates_banani_review}. In this regard, an interesting analysis was carried out by Saleh et al. in \cite{saleh}, where a system of two different kinds of nanostars with interspecies linkers was studied experimentally, highlighting physical trends (especially in connection with interspecies surface tension variations with the concentration of cross-linkers) which we reproduce here.

\section{Methods}

\subsubsection{Cahn-Hilliard equation}

The Cahn-Hilliard (CH) equation is a partial differential equation, derived from the assumption that the total free energy of a homogeneous system described by the number densities $\rho_i$ of the $N_s$ components at volume $V$ and temperature $T$ can be approximated as~\cite{ch}

\begin{equation}
F(T, V) = \int_V \left[ f(\{\rho_i\}) + \frac{K}{2} \sum_{i=1}^{N_s} (\vec \nabla \rho_i)^2 \right] dV,
\label{eq:F_tot}
\end{equation}

\noindent
where $f(\{\rho_i\})$ is the Helmholtz free-energy density, $\{\rho_i\}$ is the set of order parameters relevant to the system, and $K$ is a coefficient linked to the free energy penalty that comes with the creation of an interface between two phases. To simplify the model, we neglect cross-species interfacial terms and we implicitly set $K$ to be the same for all species.

By connecting the spatial variation of the order parameters with the spatial variation of a ``generalised'' chemical potential, it is possible to obtain a continuity equation that expresses locally the conservation of the total mass of the system~\cite{ch}:

\begin{equation} \label{eq:cahn-hilliard}
\frac{\partial \rho_j}{\partial t} = M \nabla ^{2} \left(\frac{\partial f(\{\rho_i\})}{\partial \rho_j} - K \nabla ^{2} \rho_j \right).
\end{equation}

\noindent
 In Eq.~\eqref{eq:cahn-hilliard} $t$ is the time, $M$ is a positive mobility coefficient setting the time scale of the system  which we assume to be density-independent, $\nabla^2$ is the spatial Laplacian operator, and the terms in brackets correspond to the generalised chemical potential of species $j$, $\mu_j = \partial f(\{\rho_i\})/\partial \rho_j - K \nabla^2 \rho_j$. This is the CH equation for the $j$-th number density which, given initial conditions and appropriate boundary conditions, can be numerically integrated in time to obtain the evolution of the system.

We discretise Eq.~\ref{eq:cahn-hilliard} in time with time step $\Delta t$, and in space by using bins of linear size $\Delta x$. We consider periodic boundary conditions. In order to integrate Eq.~\eqref{eq:cahn-hilliard-rescaled} in time we have implemented one explicit and two semi-implicit methods. 
Indeed, in addition to the explicit Euler method, which we have used throughout the paper, we have also implemented the implicit-explicit Euler method~\cite{soares2023exponential}, which is solved in Fourier space, and the finite-volume scheme of Bailo \textit{et al.}~\cite{noneuler2}, discussed in  Appendix~\ref{app:numerics}. We perform simulations in one- and two-dimensions using a parallel code that runs on GPUs to improve performance~\cite{lorenzo_rovigatti_2024_14576360}. In our implementation, we use dimensionless free energies, multiplying their expressions by $\beta=\frac{1}{k_B \: T}$ for convenience. As a result, we use a rescaled mobility coefficient $M' = k_B T M$, which we set to a constant value $M'=1$ (nm s)$^{-1}$, since its value affects the speed of the computations, but not the final configuration features. Therefore, in our implementation the equation we integrate becomes

\begin{equation}
\label{eq:cahn-hilliard-rescaled}
\frac{\partial \rho_j}{\partial t} = M' \nabla ^{2} \left( \frac{\partial \beta f(\{\rho_i\})}{\partial \rho_j} - \beta K \nabla ^{2} \rho_j \right).
\end{equation}

Additional details on the numerical methods, and in particular on the integration algorithm, can be found in Appendix~\ref{app:numerics}.

\subsubsection{Wertheim free energy}\label{puresystem}

The physics of the system is contained in the expression for the free energy density $f(\{\rho_i\})$, which has to be chosen carefully in order to reproduce with satisfactory accuracy the properties of the target system. For this purpose we use Wertheim theory~\cite{wertheim1984fluids,wertheim1986fluids} for self-assembly, a mean-field theory suitably adapted to a system of DNA nanostars in a NaCl solution at fixed temperature $T$ and salt concentration $[Na^+]$. The two main assumptions of the theory are the absence of loop structures in finite-size clusters, and of double bonds between pairs of nanostars.

In the Wertheim theory, the free energy is written as a sum of two terms~\cite{lorenzo}:

\begin{equation} \label{eq:wertheim}
f(\{\rho_i\})=f_\text{ref}(\{\rho_i\}) + f_\text{bond}(\{\rho_i\}),
\end{equation}

\noindent
where $f_\text{ref}(\{\rho_i\})$ is the free energy of a system where no bonding is possible, and therefore accounts for the purely repulsive forces (mainly excluded volume and electrostatic effects acting between the negatively charged backbones of DNA strands), and $f_\text{bond}(\{\rho_i\})$ is an attractive contribution, discussed later on, stemming from the hybridisation between the complementary sticky ends of the nanostars.

Since we consider rather dilute systems, we approximate the reference free energy with a second-order virial expansion, as proposed in Ref.~\cite{lorenzo}, giving:

\begin{equation}
\label{eq:f_ref}
\beta f_\text{ref} = \rho \: \log(v_0 \rho) - \rho + B_2 \: \rho^2 + \sum_{j=1}^{N_s} \rho_j \: \log\left(\frac{\rho_j}{\rho}\right),
\end{equation}

\noindent
where $\rho = \sum_j^{N_s} \rho_j$ is the total number density, $N_s$ is the number of species, $B_2$ is the second-order virial coefficient of the nanostars (with a non-bonding sticky sequence), which has the dimensions of a volume, and $v_0$ is the inverse of the partition function of a single nanostar multiplied by the volume of the system ~\cite{mcquarrie2000statistical}; the latter can be assumed to be independent of density, so that its value does not affect the phase equilibrium. In the above expression we assume that the value of $B_2$ is the same for all pairs of species of interacting nanostars, which is an accurate approximation since the nanostars we consider have the same geometry, differing only in the sequence of the sticky ends. In the following we set $B_2 = 2190$ nm$^3$, a value that has been computed in Ref.~\cite{lorenzo} through two-body coarse-grained simulations and was shown to be weakly dependent on $T$ for the salt concentration we will be using ($0.5$ M).

The attractive part $f_b(\{\rho_i\})$, described in detail below, takes into account the contribution due to the self-assembly of the nanostars. 

\begin{figure}[h!]
    \centering
    \includegraphics[width=10cm]{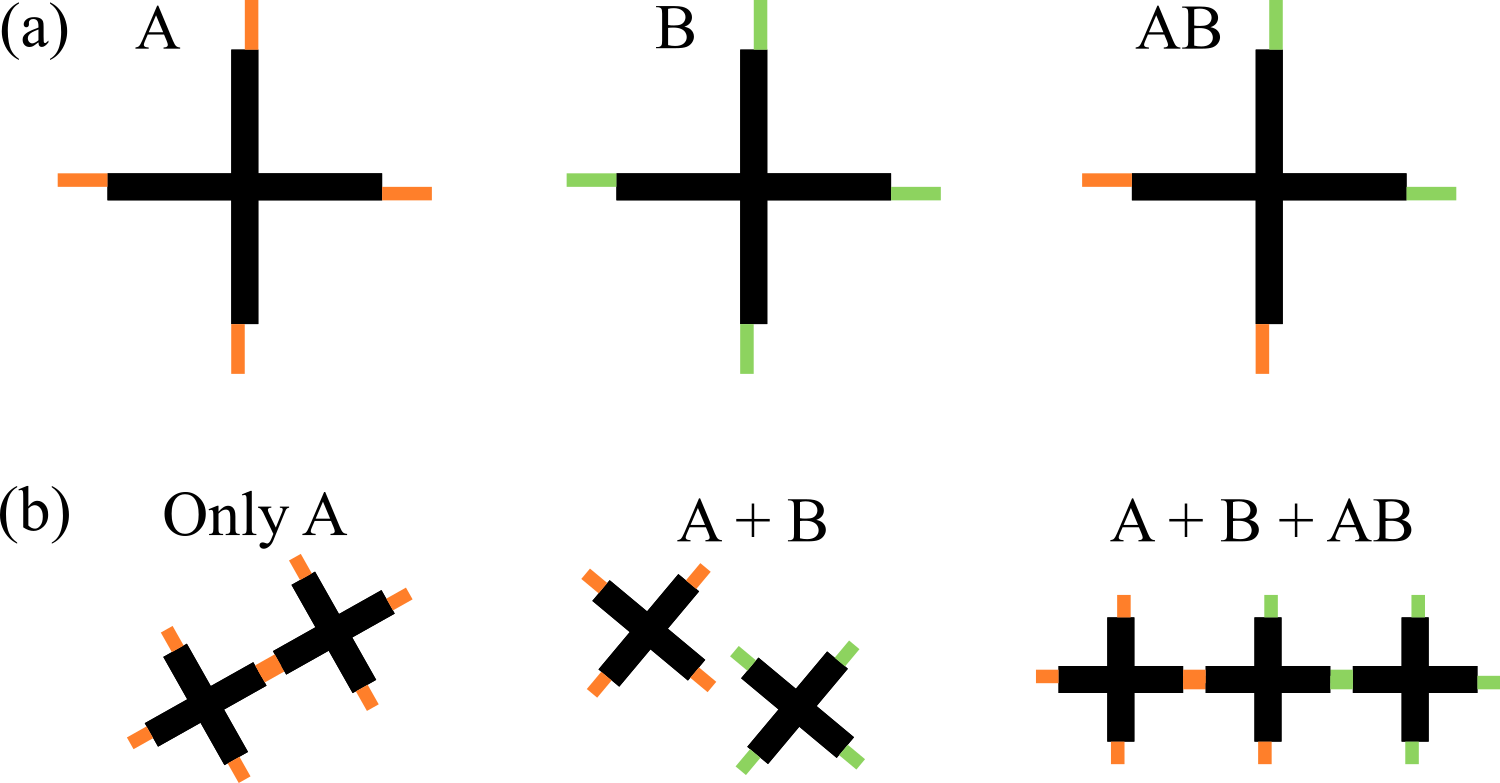}
    \caption{A sketch representing (a) the different species of DNA nanostars we consider and (b) the way they can bond: only like-colour sticky ends can bind to each other. Each nanostar is composed by $\mathcal{V}=4$ hybridised single strands of DNA (the black sections), each terminating with a single-stranded, self-complementary sticky sequence for connection with other nanostars (the coloured sections). Note that "only $A$" and "only $B$" systems have the same qualitative behaviour, while the presence of AB nanostars creates links between the A and the B particles.}
    \label{saleh_img}
\end{figure}

As an example of a phase-field description of  DNA-particle in solution we focus on the experimental system investigated by Saleh et al~\cite{saleh}, which introduced three different nanostars, $A$, $B$ and $AB$, with valence $\mathcal{V}_A = \mathcal{V}_B = \mathcal{V}_{AB} = \mathcal{V}=4$. The two species of nanostars $A$ and $B$ have respectively four $\alpha$ and $\beta$ orthogonal (\textit{i.e.} non-complementary) palindromic sticky ends each, and are combined with special cross-linker $AB$ nanostars for interspecies bonding.  These cross-linkers are designed in such a way so as to contain two sticky ends of type $\alpha$ and two sticky ends of type $\beta$, thus promoting connections between the pure A and pure B  species.  In the presence of the AB nanostars, the A particles can bind among themselves and with the AB ones, and similarly the B particles can bind among themselves and with the AB ones, thus generating connections between the A and the B particles via the AB ones.  For a graphical description of the bonding scheme, see Figure~\ref{saleh_img}.

In the following we consider systems made of only $A$ nanostars, as well as the Saleh et al~\cite{saleh} ternary mixtures of $A$, $B$, and $AB$. In the former case, the Wertheim bonding free-energy density is given by~\cite{lorenzo}

\begin{equation}
\label{eq:f_b}
\beta f_{b} = \rho_A \: \mathcal{V} \: \left( \log(X_\alpha) + \frac{1-X_\alpha}{2} \right),
\end{equation}

\noindent
where $X_\alpha(T, \rho_A, [Na^+])=\frac{-1+\sqrt{1 + 4 \Delta_{\alpha} \mathcal{V} \rho_A}}{2 \Delta_{\alpha} \mathcal{V} \rho_A}$ is the fraction of unbonded sticky ends at temperature $T$, salt concentration $[Na^+]$, and density $\rho_A$, and $\Delta_\alpha$ has the dimensions of a volume and quantifies the strength of the interaction acting between the specific complementary sequences of the sticky ends and is equal to:

\begin{equation}
\label{eq5}
\Delta_\alpha(T, [Na^+]) = v_b \: e^{-\beta \Delta G_\alpha} = v_b \: e^{-\beta ( \Delta H_\alpha - T \Delta S_\alpha)},
\end{equation}

\noindent
where $v_b=1.66$ nm$^3$ is the standard bonding volume associated with each particle, while $\Delta G_\alpha$ is the Gibbs free energy variation associated with the hybridisation of two sticky ends $\alpha$. The latter variation consists of two contributions, an enthalpic term, $\Delta H_\alpha$, and an entropic term, $\Delta S_\alpha = \Delta S^\text{salt} + \Delta S^\text{no salt}_\alpha$. The values of these two terms can be worked out from the specific sticky-end sequences of DNA bases using SantaLucia model~\cite{santa_lucia1, santa_lucia2}. Table~\ref{tab1} contains the sequences (taken from~\cite{saleh}) and free-energy parameters of the sticky ends used in this work. Note that the chosen sequences are \textit{palindromic} (i.e. self-complementary) to allow for $\alpha\alpha$ and $\beta\beta$ bonding, and that we fix the value of the salt concentration $[Na^+] = 0.5$ M throughout the whole paper.

In order to compare with numerical results, we use the one-component Wertheim free energy to calculate the phase diagram. The densities of the coexisting phases are found by using the Maxwell construction at fixed temperature~\cite{lorenzo}.

\begin{table}[H]
\begin{center}
\begin{tabular}{ |c|c|c|c|c| }
 \hline
 \multicolumn{5}{|c|}{Sticky sequences and descriptive parameters}\\
 \hline
 Sticky end & Sequence & $\Delta H [\frac{\rm cal}{\rm mol}]$ & $\Delta S_\text{salt} [\frac{\rm cal}{\rm mol K}]$ & $\Delta S_\text{nosalt} [\frac{\rm cal}{\rm mol K}]$\\
 \hline
 $\alpha$ & CGATCG & $-42200$ & $1.84 \log([Na^+])$ &  $-119.1$\\
 $\beta$ & GAGCTC & $-42400$ & $1.84 \log([Na^+])$ & $-120.6$\\
 \hline
\end{tabular}
\end{center}
\captionof{table}{Nucleotide sequences for the two kinds of sticky ends considered in this study, and enthalpic and entropic variations associated with their hybridisation calculated according to Ref.~\cite{santa_lucia1, santa_lucia2}.}
\label{tab1}
\end{table}

With the free energy contributions given in Eqs.~\ref{eq:f_ref} and~\ref{eq:f_b}, the Cahn-Hilliard equation can be written explicitly for this pure system, giving:
\begin{equation}\label{ch_eq2}
\frac{\partial \rho_A}{\partial t} = M' \nabla ^{2} \left 
 (\beta \mu_\text{ref} + \beta \mu_b + \beta \mu_{int} \right ),
\end{equation}

\noindent
in which the chemical potential contributions from the reference, bonding and interface terms come respectively in the form $\beta \mu_\text{ref}(\rho_A)=\frac{df_\text{ref}(\rho_A)}{d\rho_A} = \log(\rho_A)+2 B_2  \rho_A$, $\beta \mu_{b}(\rho_A) = \frac{df_b(\rho_A)}{d\rho_A} = \mathcal{V} \log(X_\alpha(T, \rho_A))$ and $\beta \mu_\text{int}(\nabla \rho_A)= - \beta K \nabla^2 \rho_A$.

For Saleh's mixture, the attractive contribution $f_b$ takes the following form~\cite{lorenzo}:

\begin{equation}
\beta f_b = \rho_A \mathcal{V}\left(\log(X_\alpha) - \frac{X_\alpha}{2} + \frac{1}{2}\right) + \rho_B \mathcal{V} \left(\log(X_\beta) - \frac{X_\beta}{2} + \frac{1}{2}\right) + \rho_{AB} \frac{\mathcal{V}}{2} \left(\log(X_\beta) - \frac{X_\beta}{2} + \log(X_\alpha) -\frac{X_\alpha}{2}+ 1\right),
\end{equation}

\noindent
where $X_\alpha$ and $X_\beta$ correspond to the fraction of unbonded sticky ends of type $\alpha$ and $\beta$, respectively, and are given by:

\begin{equation}
X_\gamma=\frac{-1+\sqrt{1+4 c_\gamma \Delta_\gamma}}{2 c_\gamma \Delta_\gamma},
\end{equation}

\noindent
where $\gamma \in \{\alpha, \beta\}$ and $c_\gamma$ is the number density of sticky ends of type $\gamma$, \textit{i.e.} $c_\alpha = \mathcal{V}(\rho_A + \frac{\rho_{AB}}{2})$ and $c_\beta = \mathcal{V} (\rho_B + \frac{\rho_{AB}}{2})$.

Finally, the system of coupled differential equations to be solved to obtain the time evolution dynamics of the three density fields can be written as:

$$
\frac{\partial \rho_i}{\partial t} = M' \nabla^2 \left( \beta \mu^i_\text{ref} + \beta \mu^i_{b} + \beta \mu^i_{int}\right), \qquad i \in \{A, B, AB\}
$$

where the chemical potential contributions for species $i$ amount to $\beta \mu^i_\text{ref}(\rho_i, \rho)=2B_2\rho + \log \rho_i$, $\beta \mu^i_b(\{\rho_j\})=\sum_{\gamma \in \Gamma(i)} \log X_{\gamma}$ and $\beta \mu^i_\text{int}(\nabla \rho_i) = -\beta K \nabla^2 \rho_i$. Here $\Gamma(i)$ refers to the set of sticky ends present on nanostars of type $i$, and $K$ is related to the interfacial cost, which we assume independent of $i$.

\section{Results}

\subsection{Pure systems}

We use the one-component system made by $A$ nanostars to assess the effectiveness and reliability of the present Cahn-Hilliard based approach, as well as to select an optimal value for the $\beta K$ constant.

\begin{figure}[h!]
    \centering
    \includegraphics[width=14cm]{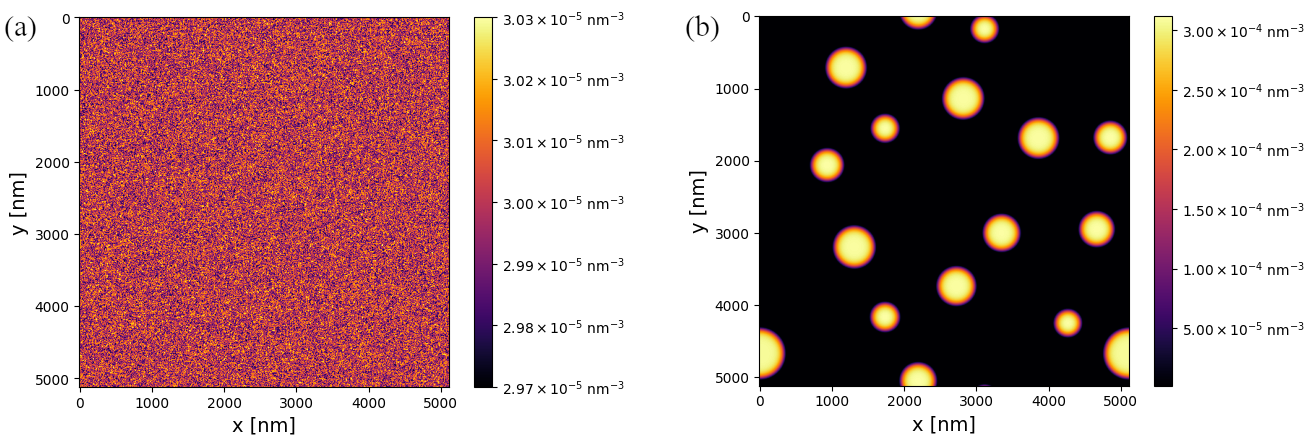}
    \caption{Initial and final configurations of a simulation of a one-component $A$ system at $T = 303.15$ K. (a) Initially the system is initialised in a quasi-homogeneous configuration, where each bin has a nanostar number density close to $\rho_A = 3 \times 10^{-5}$ nm$^{-3}$. (b) After $\approx 10^8$ steps run with $\Delta t = 10^{-5}$ s, the system is well separated, with liquid droplets coexisting with a gas background.}
    \label{liquidgas}
\end{figure}

It was shown experimentally that, under the right thermal and density conditions the only-A system phase separates into a high-density phase containing a network of bonded nanostars (the liquid), and a low-density phase composed of mostly unbonded nanostars (the gas)~\cite{biffi}. Within the present approach, we also find that any initial configuration defined by 
small random density fluctuations superimposed to an average value 
evolves according to the CH equations or toward a completely homogeneous system, or toward a phase-separated system, as shown in Figure~\ref{liquidgas} for a 2D system. 

The coefficient $K$ appearing in the Cahn-Hilliard equation is associated to the Helmholtz free energy cost associated with the creation of an interface between two different phases. Indeed, the free energy penalty for interface creation comes in the form $\frac{K}{2}|\mathbf{\nabla} \rho|^2$, which means that for a given interface cost, increasing the value of $K$ decreases the density gradient magnitude $|\mathbf {\nabla} \rho|$, thus making the interface smoother and wider. Of course, decreasing $K$ does the opposite.

Here we set the value of $\beta K$ by comparing the theoretical gas-liquid phase diagram, obtained by using the Maxwell construction with the free energy given by Eq.~\eqref{eq:wertheim}, with numerical results of 1D Cahn-Hilliard simulations. We fix the timestep, $\Delta t = 10^{-6}$ s, and the grid spacing $\Delta x = 10$ nm, and run simulations at four different temperatures, $T = 298.15$, $303.15$, $308.15$ and $309.15$ K, for $10^{10}$ time steps. Note that the highest $T$ is just above the theoretical critical temperature $T_c=308.44$ K, and therefore does not phase separate. For each state point we use three values of the interface constant, $\beta K = 10^5$, $10^6$ and $10^7$ nm$^5$, and two different initial conditions: \textit{separated} and \textit{homogeneous}. The former refers to a starting configuration where the two halves of the system are initialised with the theoretical densities of the gas and the liquid, $\rho_g$ and $\rho_l$, connected by sigmoidal curves. By contrast, in \textit{homogeneous} simulations we start with a system of average density $(\rho_g + \rho_l) / 2$, with random fluctuations around this value.

\begin{figure}[h!]
    \centering
    \includegraphics[width=0.45\linewidth]{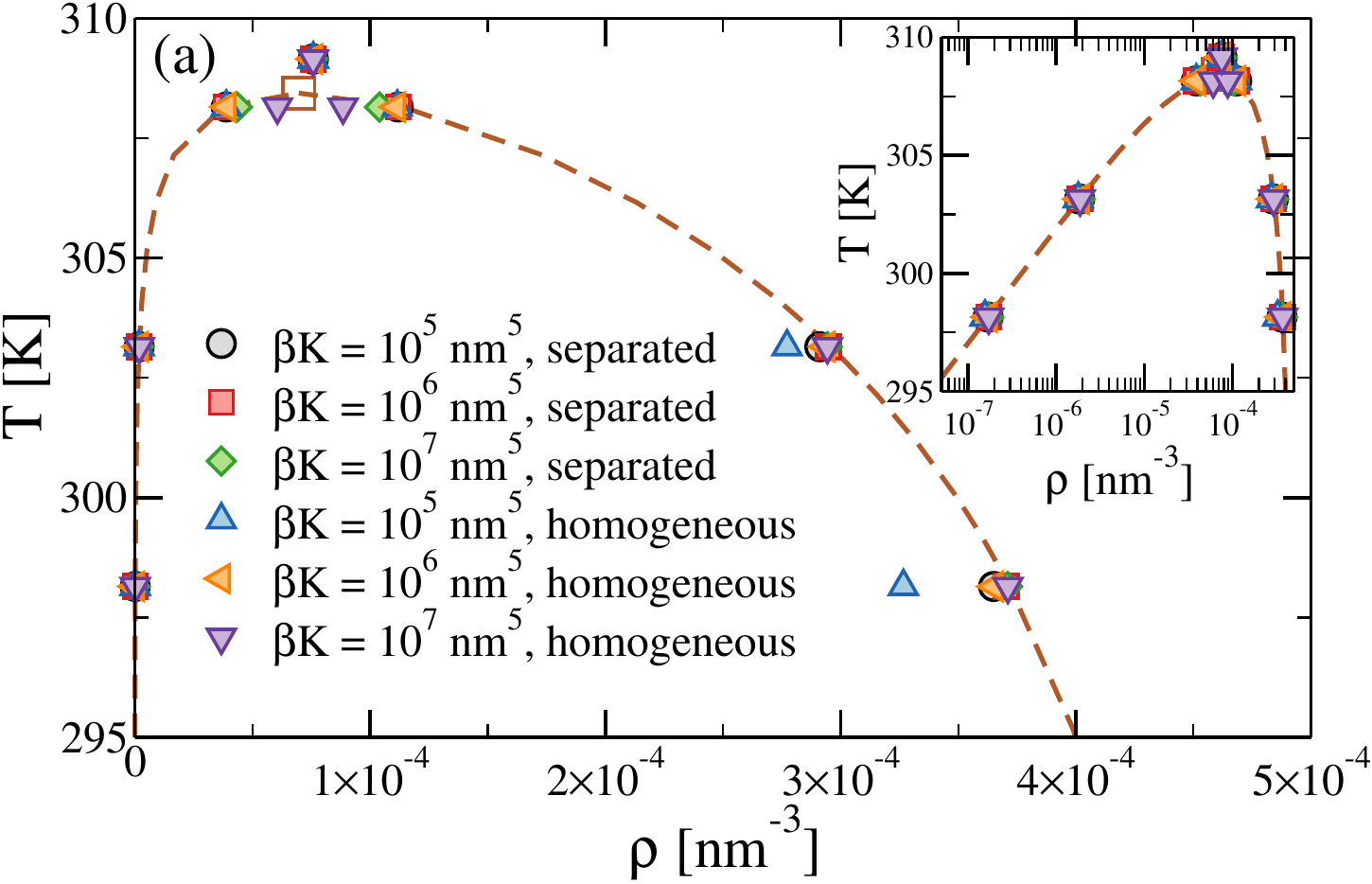}
    \includegraphics[width=0.48\linewidth]{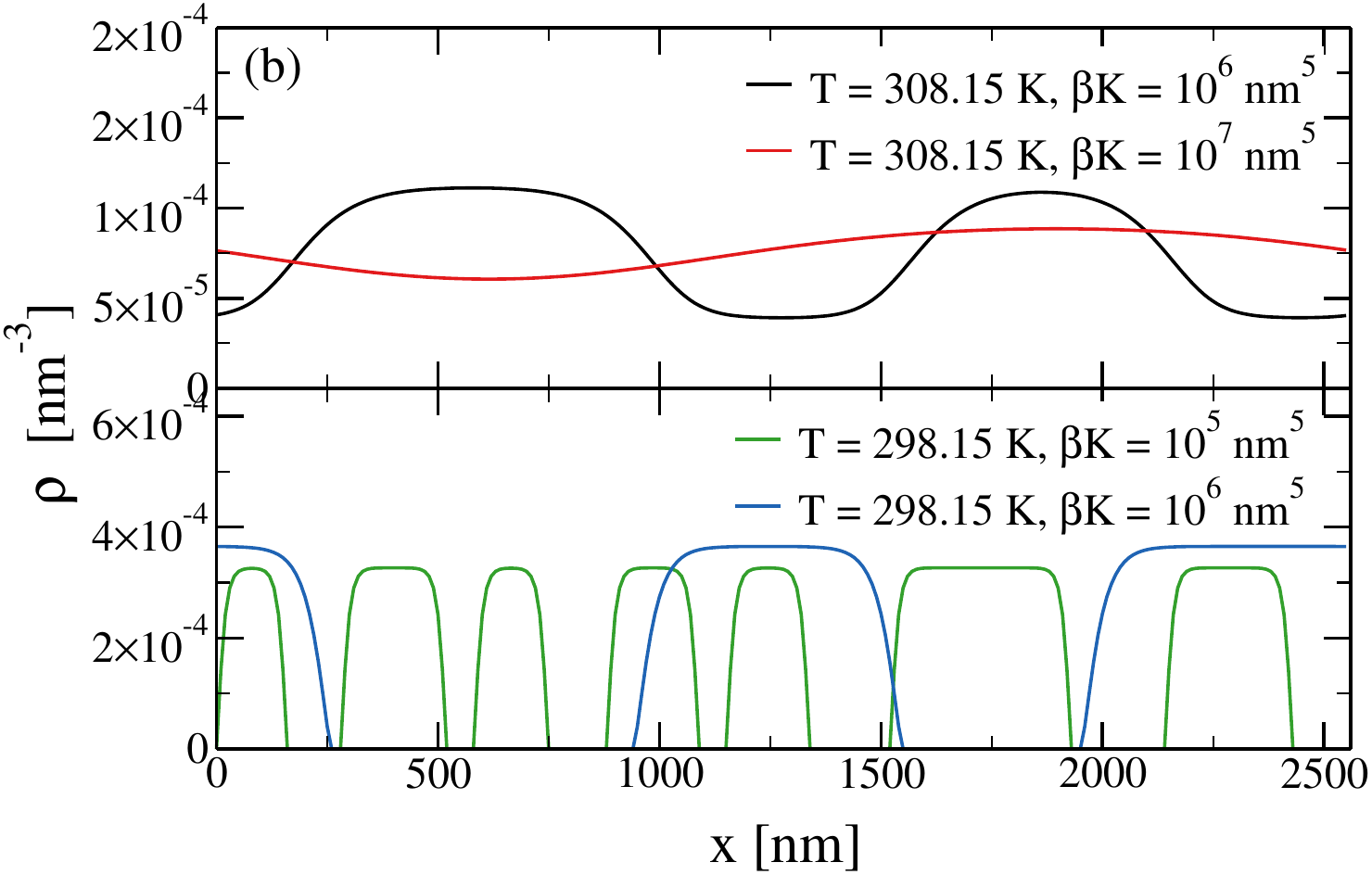}
    \caption{(a) A comparison between the theoretical (brown dashed lines) and numerical (filled symbols) phase diagrams, for three values of $\beta K$ and two different initial conditions. The brown open square signals the position of the critical point $(\rho_c=6.67 \times 10^{-5}$ nm$^{-3}, T_c=308.44 $ K$)$. Inset: the same data is shown with a logarithmic scale for $\rho$ to highlight the agreement at low density. (b) Final density profiles for simulations run at (top) $T = 308.15$ K and (bottom) $T = 298.15$ K, for two different values of the interfacial cost $\beta K$.}
    \label{fig:pd}
\end{figure}

Figure~\ref{fig:pd}(a) shows the comparison between theory and simulation. First of all, we note that all simulations predict the homogeneity at the highest temperature considered, in agreement with theory. Secondly, below the critical temperature $T_c$, the densities of the two coexisting phases are well-reproduced by the simulations, with two exceptions. For $T = 308.15$ K, which is just below $T_c$, the $\beta K = 10^7$ nm$^5$ coexisting densities for the initially homogeneous systems are closer than what they should be. We identify this disagreement as a finite-size effect, as at this temperature and value of $\beta K$ the width of the interface becomes comparable with the box size. By contrast, if $\beta K = 10^5$ nm$^5$, low-temperature simulations started from homogeneous configurations tend to underestimate the liquid density. This happens because the dynamics slows down massively, as seen by the much smaller values of the time derivative of the density field for the $\beta K = 10^5$ nm$^5$ system (not shown), which therefore remains stuck in a metastable configuration. The final density profiles of systems displaying these two effects are shown in Figure~\ref{fig:pd}(b) and compared to density profiles of systems with $\beta K = 10^6$ nm$^5$, which achieve the correct liquid density at the end of the simulation. 

These results show that setting $\beta K = 10^6$ nm$^5$ provides, for the system size studied in this study, 
correct estimates of the equilibrium densities  and that 
equilibration takes place in reasonable computational times for the explored range of temperatures. Thus we decided to fix $\beta K$ to this value for the rest of this work.

We also estimate the surface tension $\gamma$ associated to the gas-liquid interface, which is defined as the free-energy cost per unit area of forming the interface. Since we have access to the total free energy of a system through Eq.~\eqref{eq:F_tot}, we can directly estimate $\gamma$ by using its definition. In order to do so, we evaluate the free energy of two systems initialised with the coexisting densities of either the gas or the liquid phase at temperature $T$, yielding $F_\text{gas}$ and $F_\text{liquid}$. We then put these two systems in contact and simulate the formation of two interfaces (through periodic-boundary conditions) of interface area $A$. At long times, the free energy of the system converges to a value $F_\text{coex}$. The surface tension for each value of T can then be estimated as

\begin{equation}
    \gamma = \frac{F_\text{coex} - (F_\text{gas} + F_\text{liquid})}{2 A},
    \label{eq:gamma}
\end{equation}

In 1D we use systems made of $N = 128$ bins, and $A = \Delta x^2$, since the system can be thought of as a parallelepiped of physical dimensions $N\Delta x \times \Delta x \times \Delta x$. By contrast, in 2D $A = N \Delta x^2$, since the size of the simulation box is $N\Delta x \times N \Delta x \times \Delta x$.

\begin{figure}[!h]
    \centering
    \includegraphics[width=0.45\linewidth]{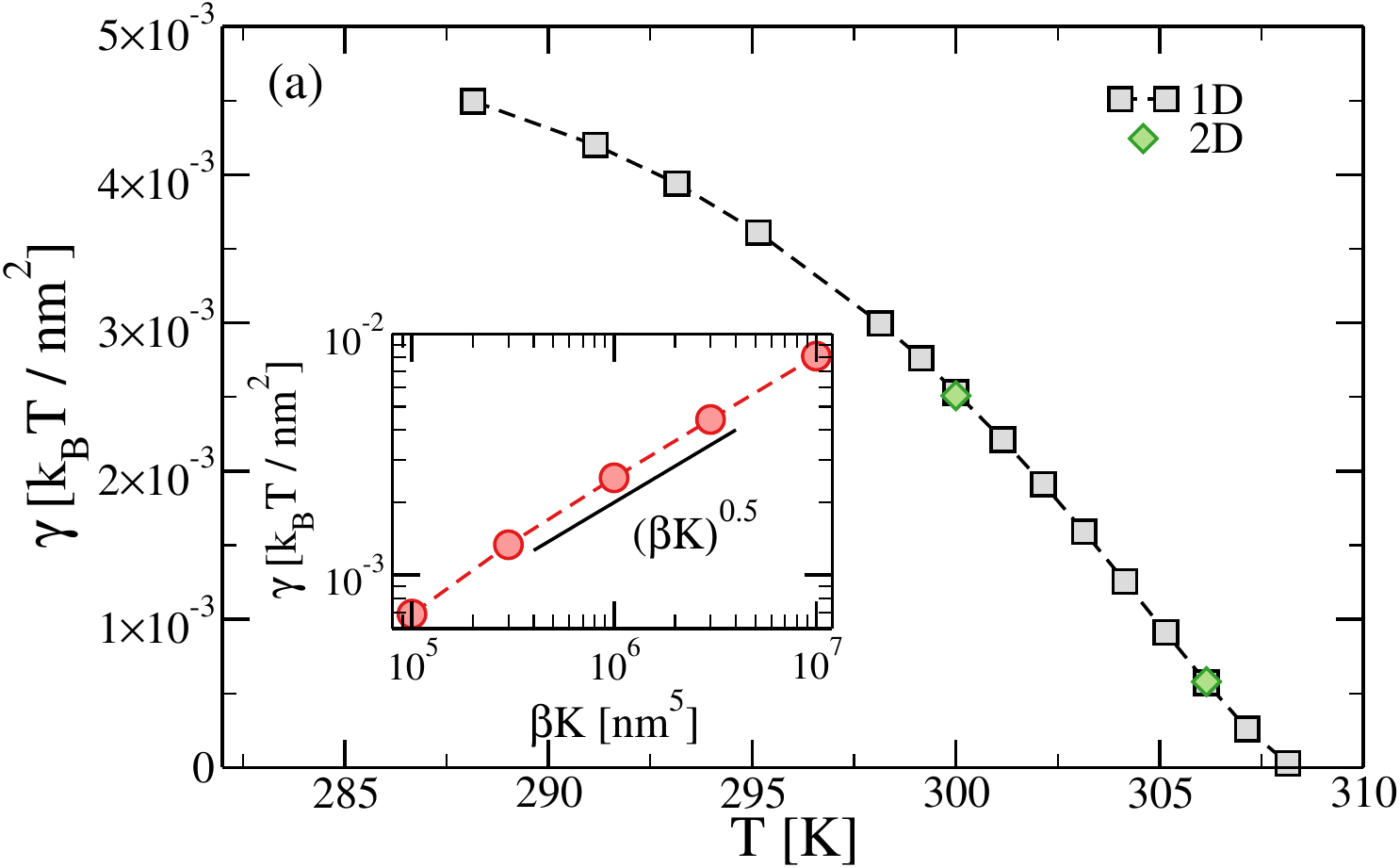}
    \includegraphics[width=0.45\linewidth]{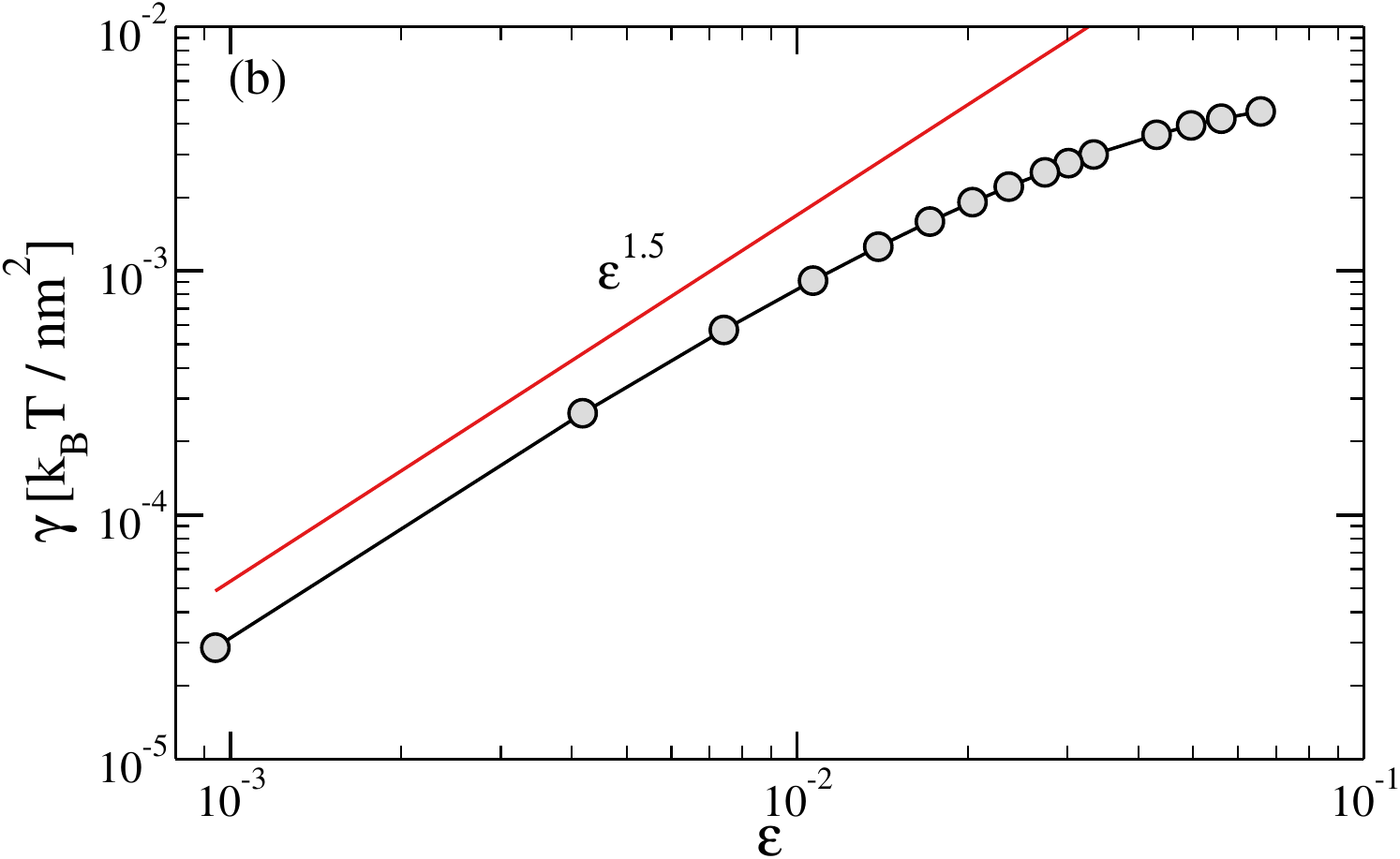}
    \caption{(a) The surface tension of the one-component $A$ system, $\gamma$, as a function of temperature, calculated by putting in contact the two phases at coexistence and waiting for the correct interface to develop. The black and green symbols refer to 1D and 2D systems, respectively. Inset: $\gamma$ as a function of $\beta K$ for $T = 300$ K. The observed dependence is compatible with a square root. (b) The surface tension as a function of $\varepsilon = (T_c - T) / T_c$, the scaled distance from the critical temperature. Close to the critical point  (\textit{i.e.} for $\epsilon < 10^{-2}$) the observed dependence on $\varepsilon$ is compatible with a power-law with exponent $1.5$, as expected for a mean-field theory.}
    \label{fig:gamma_T}
\end{figure}

As shown in Figure~\ref{fig:gamma_T}(a), the surface tension increases upon decreasing temperature, with the dependence on $T$ weakening as the system cools down. Such a dependence on $T$ is due to the parameters that control the density of the coexisting phases, which in the theory we use are the excluded volume $B_2$ (for the range of temperatures studied we consider the latter to be $T$-independent, see~\cite{lorenzo}) and the attraction volume $\Delta(T)$. At lower values of $T$, $\Delta$ increases steeply, resulting in an effective increase of the attractive force between the nanostars, which leads to a larger gap between the coexistence densities $|\rho_{\text{liquid}}-\rho_{\text{gas}}|$ and to steeper interfaces, both of which result in a higher liquid-gas surface tension (for an explanation of this fact, see Appendix~\ref{app:gamma}). Interestingly, sufficiently apart (1-2 K) from the critical point, $\gamma \sim 10^{-3} \; k_BT/\text{nm}^2$, which is compatible with experimental estimates of the surface tension of DNA nanostars~\cite{jeon2018salt,saleh}. This agreement validates the value of $\beta K$ that was chosen for numerical convenience. We also computed the surface tension for two 2D systems using the GPU code, and the results are fully compatible with the 1D CPU data, suggesting a lack of dependence of $\gamma$ on the system dimensionality, which we expect to hold also for 3D systems. As shown in Figure~\ref{fig:gamma_T}(b), we find that, close to the critical point (\textit{i.e.} for $\epsilon < 10^{-2}$), the surface tension is compatible with a power-law dependence with exponent $1.5$, in agreement with the mean-field nature of Wertheim theory~\cite{rowlinson2013molecular}. Away from the critical point, an analysis of experimental data based on the Van der Waals' theory of liquids suggests that the $T$-dependence of $\gamma$ is compatible with an exponential behaviour~\cite{conrad2022emulsion}. Unfortunately the temperature range in which we can simulate is too limited to test this suggestion.\\
Our estimates of $\gamma(T)$ have been further contrasted with a theoretically-derived computation of the same quantity, obtained through free energy minimisation techniques from an initially phase-separated liquid-gas system modelled by a $\tanh(\alpha x)$ density profile function. The detailed procedure followed for this calculation, as well as the results of the comparison with CH-derived data, are described in Appendix~\ref{app:gamma}.\\
Finally, in the inset of Figure~\ref{fig:gamma_T}(a) we show that $\gamma \propto \sqrt{K}$, as expected in CH simulations~\cite{cahn1958free}.\\
\subsection{Three-component systems}

\begin{figure}
    \centering
        \subfloat[\label{fig_theoretical}]{\includegraphics[width=0.7\linewidth]{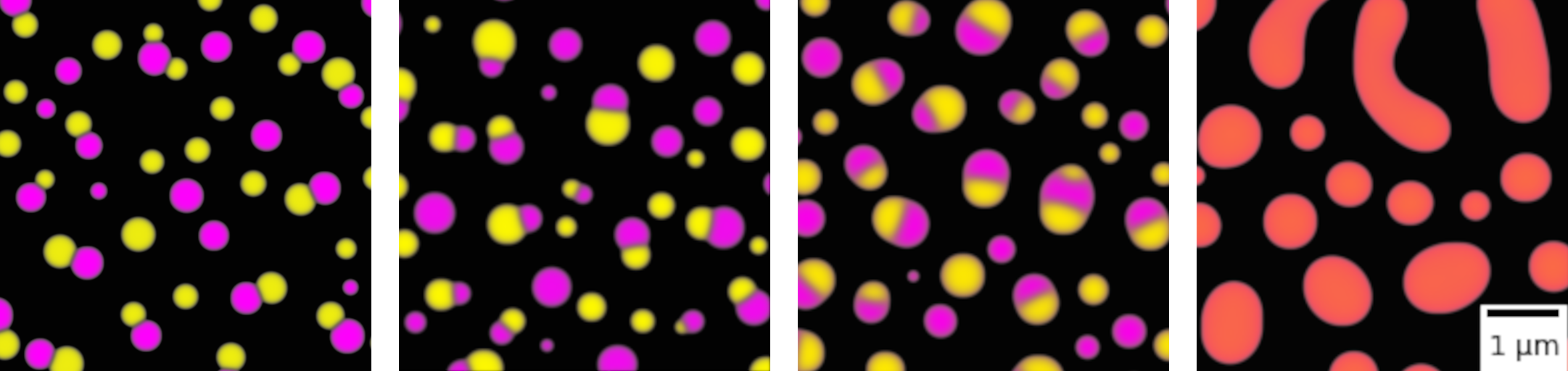}}\\
        \subfloat[\label{fig_linkers}]{\raisebox{0cm}{\hspace{1.6cm}\includegraphics[width=0.79\linewidth]{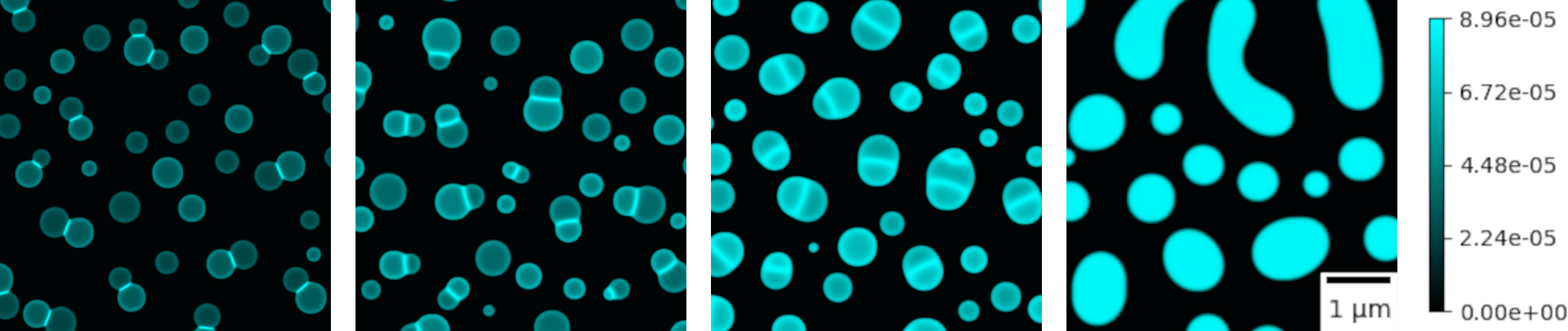}}}\\
        \subfloat[\label{fig_experimental}]{\includegraphics[width=0.7\linewidth]{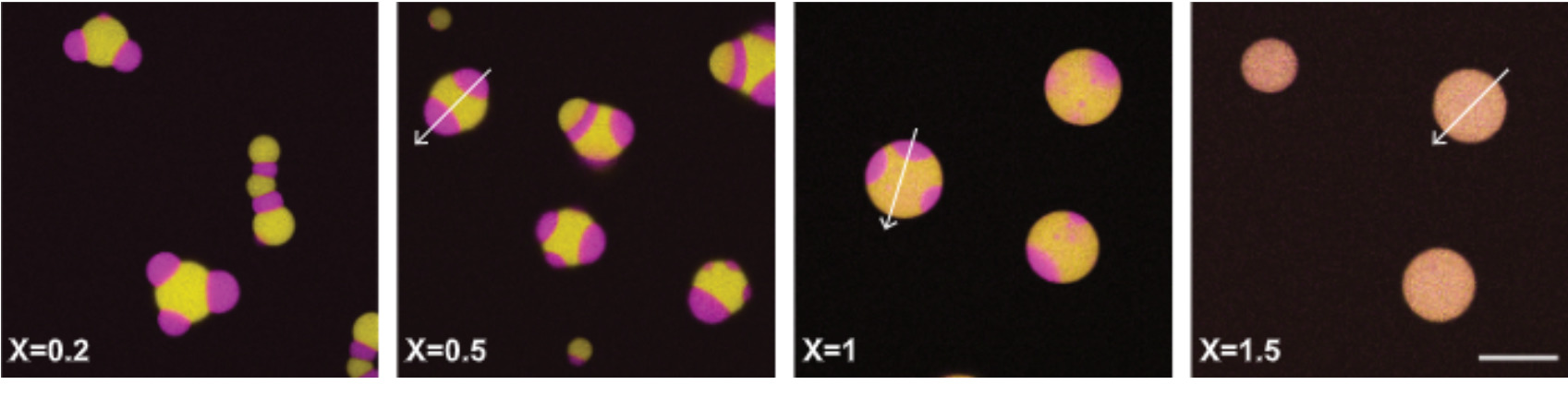}}

    \caption{(a) Simulated configurations of phase-separated systems as the cross-linker fraction $c_X$ is increased, from left to right. Each  pixel is coloured according to the density of each species it contains, where species $A$ and $B$ are associated to magenta and yellow, respectively. The simulations have been run at $T = 300$ K and average
densities $\rho_A = \rho_B = 2 \cdot 10^{-5}$ nm$^{-3}$ and $\rho_{AB} = c_X \rho_A$, with $c_X \in \{0.2, 0.5, 1.0, 1.5\}$. The images displayed here are the bottom-right fourths (size $512\times 512$ pixels) of the originally simulated system (size $1024 \times 1024$ pixels). The scale bar corresponds to $1$ $\mu$m.
    (b) The density of cross-linkers for the same configurations shown in panel (a). Note that a colour bar linking pixel intensity to the cross-linker number density in [nm]$^{-3}$ is given; a similar approach is not possible for panel (a), where each pixel colour is the sum of individual components corresponding to the concentrations of the $A$ and $B$ nanostars. (c) Experimental configurations of phase-separated systems corresponding to increasingly higher (from left to right) values of $c_X \in \{0.2, 0.5, 1.0, 1.5\}$, taken at $T=293.15$ K and $\rho_A=\rho_B=3.0 \times 10^{-6}$ nm$^{-3}$. Species $A$ and $B$ are tagged with two different fluorescent molecules: magenta and yellow, respectively. The scale bar corresponds to $20$ $\mu$m. Adapted with permission from Ref.~\cite{saleh}. Copyright 2020 American Chemical Society.}
    \label{configurations}
\end{figure}

We now move on to the computational study of the three-component system of Saleh \textit{et al}, who investigated experimentally the properties of phase separated systems of $A$ and $B$ nanostars for different values of the fraction of the AB cross-linkers $c_X=\frac{\rho_{AB}}{\rho_A}$ in ~\cite{saleh}; their study was carried out by quenching homogeneous mixtures of nanostars in the binodal region at fixed $c_X$, and letting the system phase separate, monitoring the concentration of the $A$ and $B$ species by fluorescence microscopy. In the simulations for this part, we fix the temperature to $T = 300$ K (which is below the critical temperature $T_c$, otherwise the system would remain homogeneous according to Wertheim theory) and simulate grids made of $1024 \times 1024$ bins of width $\Delta x = 10$ nm. Movies showing the evolution of the system from an initially homogeneous configuration to a phase separated state can be found in the Supplementary Material. The selected temperature is slightly higher than the experimental one, $T = 293.15$ K. While this difference is not expecting to change the physics of the phase separation process, it allows us to follow the phase separation kinetics for longer times, helping the comparison with the experimental results.  We also note 
that since Wertheim theory is known to reproduce the thermodynamics of DNA nanostars only in a semi-quantitative manner~\cite{lorenzo,locatelli2018accurate,conrad2022emulsion}, it is not obvious that using the experimental temperature would result in a more precise comparison.

We focus on the effect of the cross-linkers by simulating systems with a fixed and equal average density of $A$ and $B$ nanostars, \textit{i.e.} $\rho_A = \rho_B$, to which we add a fraction $c_X \rho_A$ of cross-linkers. When $c_X$ is zero or small, then both the A and the B nanostars undergo a gas-liquid phase separation similar to the one-component case, where the $A$ and $B$ species form completely demixed liquid droplets. As $c_X$ increases, the tendency of the cross-linkers to bond to both species lowers the $A-B$ liquid-liquid surface tension, leading to systems made of fused droplets, which are still partially demixed. For $c_X > 1$, the abundance of cross-linkers requires that $A$ and $B$ co-localise, resulting in liquid droplets where the $A$ and $B$ species are fully mixed. As shown in Figure~\ref{fig_theoretical}, the solutions of the CH equation for the chosen values of $c_X$ display exactly this behaviour, in qualitative agreement with the experimental study results reproduced in Figure~\ref{fig_experimental} [22], as well as with recent experiments on RNA-based condensates [34]; however,
it is worth noting that minor differences occur. For example, at high linker fraction $c_x=1.5$, the shape of the simulated liquid droplets looks elongated with respect to the (expected) rounded configurations seen in experiments; we ascribe this difference to the increased time it takes for $c_X=1.5$ to reach the final stage of coarsening. Also note that the average density of the numerical and experimental systems differ by almost an order of magnitude, owing to the different (smaller) time scales accessible with our method.

In Figure~\ref{fig_linkers} we also show the density of cross-linkers: at small values of $c_X$ they are concentrated at the A-B liquid interface, while as $c_X$ increases, they start accumulating also in the bulk phases, up to complete mixing for $c_X=1.5$. We note that the total density of the droplets decreases as $c_X$ increases. This decrease could be interpreted as a result of an effective decrease of the valence inside the droplets, which are mainly composed by a tetravalent majority species, and crosslinkers which, when isolated, can form only two bonds with it. As shown in~\cite{lorenzo}, a reduced valence results in a lower density of the condensed phase. It would be interesting to test this prediction in experiments. 

\begin{figure}
    \centering
    \subfloat[\label{cont_ang}]{\includegraphics[width=0.2\linewidth]{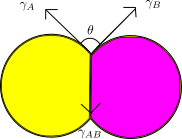}}
    \\
    \subfloat[\label{X_ang}]{\includegraphics[width=0.5\linewidth]{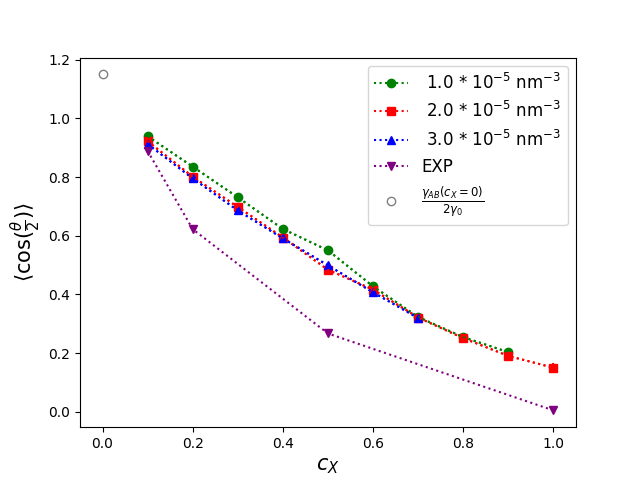}}
        
    \caption{(a) Cartoon showing the contact angle between two liquid droplets of species A and B.\break
    (b) Surface tension study as a function of the cross-linker fraction $c_X$ in comparison with experimental data. The theoretical estimates at $\rho_A=\rho_B=\{1.0 \times 10^{-5}, 2.0 \times 10^{-5}, 3.0 \times 10 ^{-5}\}$ nm$^{-3}$ have been obtained from averaging over $N=4$ equilibrated configurations at each state point; the calculation of $\cos{\theta}$ for each binary cluster was automated exploiting Python libraries by fitting curves on each species contour and calculating derivatives at the contact points. The empty point $\frac{\gamma_{AB}(c_X=0)}{2 \gamma}$ has been derived empirically by measuring $\gamma_{AB}(c_X=0)$ similarly to $\gamma$, \textit{i.e.} through total free energy subtraction methods. Note that in the figure error bars (showing the mean-squared error calculated over different droplets and different configurations for each value of $c_x$) are smaller than symbol size.}
    \label{angle_graph}
\end{figure}

Similarly to the procedure followed in the experimental paper~\cite{saleh}, from the final simulation configurations we identify all the droplets formed by two fused pure-$A$ and pure-$B$ droplets, and then measure the contact angle $\theta$ formed at each junction between the gas, liquid $A$ and liquid $B$ phases (see Figure~\ref{cont_ang}). Exploiting the von-Neumann convention to express the liquid-liquid surface tension between the A and B nanostar bulk phases in terms of $\theta$, and imposing
mechanical equilibrium, the liquid-gas and A-B surface tensions, $\gamma$ and $\gamma_{AB}$, are connected by $\gamma_{AB} = 2\gamma \cos{\frac{\theta}{2}}$. In
the experimental study, it was assumed that $\gamma_{AB} (c_X = 0) = 2\gamma$. Here we can independently estimate $\gamma_{AB} (c_X = 0)$ by using an expression analogous to Eq. 11, with $F_A$ and $F_B$ , the total free energy of two pure A and B systems, in place of $F_\text{gas}$ and $F_\text{liquid}$:
\begin{equation}
\gamma_{AB}=\frac{F_{AB} - F_{A} - F_{B}}{2 N \Delta x^2 }
\end{equation}
where $F_{AB}$ refers to the total free energy of the final equilibrated system, obtained by juxtaposing the two pure liquids A and B, and letting the interfaces form.

Figure~\ref{angle_graph} shows $\cos{\frac{\theta}{2}}$ as a function of $c_X$, for three different average densities of species $A$ and $B$, with $\rho_A=\rho_B$. All curves display a decreasing trend for surface tension as the fraction of linkers $c_X$ is increased, in agreement with theory; it is worth noticing that the average densities $\rho_A=\rho_B$ do not seem to significantly affect the plotted quantity, as suggested by the quasi-overlapping points obtained at different densities, although the low-density measurements marginally diverge from the others, which is likely due to the smaller size of clusters altering the contact angle. We note that the agreement between simulation and experiment is only semi-quantitative, as the numerical results are always higher than the experimental data, which falls off to zero for $c_x \approx 1$; this mismatch may be due to the assumptions behind the theory we use (including the approximations of Wertheim theory described in the corresponding subsection of the \textit{Methods} and the lack of thermal diffusion in the context of Cahn-Hilliard equation), to the significantly lower average density in which experiments were carried out, to the different temperature of the experiments (namely $T=293.15$ K), and to generally different and incomparable observation time scales. Unlike the experimental points, our simulation-extracted measurements also seems to describe a linear variation of $\langle\cos{\frac{\theta}{2}}\rangle$ with respect to $c_X$ at low values of $c_X$.

\section{Conclusions}

In conclusion, we have shown that the venerable Cahn-Hilliard equation can be used to study the thermodynamics of phase separation in DNA-based systems by leveraging an expression for the free energy that models the self-assembly process. We have shown that this approach can be generalised to mixtures of many molecular species, and that its theoretical consistency and qualitative agreement with experimental trends make it a powerful means of characterisation for the properties of self-assembling systems. Indeed, the phase-field method introduced here offers the possibility of simulating larger time and length scales with respect to molecular descriptions.  
Our study shows that the Wertheim theory~\cite{wertheim1984fluids,wertheim1986fluids}, associated to the CH equation provides an accurate description of available experimental data, although minor differences are still observed in the dependence of the contact angle on the fraction of cross-linkers. It will be interesting to understand in depth the origin of these differences and if these are related to the simplifying choices adopted (for example the absence of thermal noise and the hypothesis of equal $K$ for each component), or to other factors such as the different density in the experiments and in the simulations, or the different time covered in experiments and in our study.
There are few aspects that can be optimised to fully uncover the potential of this approach, including integrating thermal noise in the Cahn-Hilliard equation (as done, \textit{e.g.}, in~\cite{diffusion}) and potentially increasing the integration time-step through more refined numerical algorithms~\cite{noneuler1, noneuler2}.
We envision applications for this method in the context of multi-component liquid-liquid phase separation in cells~\cite{banani2016compositional,riback2020composition}, to investigate the kinetics of self-assembly in complex geometries~\cite{li2016multi} (with the \textit{caveat} that some form of thermal diffusion accounting for Brownian motion and coalescence effects has to integrated in the model to correctly capture the latter), or to model the coupling between thermodynamics and matter flow (\textit{e.g.} by coupling the Cahn-Hilliard equation to Stokes or Navier-Stokes equations~\cite{kim2005phase,boyer2010cahn}).

\section*{Supplementary Material}

The Supplementary Material contains two movies showing the time evolution of phase separation in an initially homogeneous ternary mixture of DNA nanostars with $[Na^+]=0.5$ M, $T=300$ K, $\rho_A=\rho_B=2.0 \cdot 10^{-5}$ nm$^{-3}$, $c_X=0.5$ and $c_X = 1.5$. The movies show the formation of tiny, densely distributed and interconnected liquid droplets of both species across the sample from an initially mixed phase, which then tend to get larger and less numerous as coarsening progresses. This behaviour is consistent with thermodynamic expectations, since at infinite time only two, possibly adhered, droplets (one per species) should be expected.

\section*{Acknowledgements}

We thank Silvio Bianchi and Omar A. Saleh for useful discussions. We acknowledge support by ICSC – Centro Nazionale di Ricerca in High Performance Computing, Big Data and Quantum Computing, funded by European Union – NextGenerationEU, and CINECA-ISCRA for HPC resources. L.R acknowledges support from MUR-PRIN Grant No. 20225NPY8P.

\appendix
\section{Additional details on the numerical algorithm}
\label{app:numerics}

In order to integrate Eq.~\eqref{eq:cahn-hilliard-rescaled} in time we have implemented one explicit and two semi-implicit methods. Indeed, in addition to the explicit Euler method, which we have used throughout the paper, we have also implemented the implicit-explicit Euler method~\cite{soares2023exponential}, which is solved in Fourier space, and the finite-volume scheme of Bailo \textit{et al.}~\cite{noneuler2}. The latter two have better properties when it comes to mass conservation and free-energy dissipation, which usually means that they make it possible to use larger values of $\Delta t$. This is especially true for the scheme of Ref.~\cite{noneuler2}, which is unconditionally stable. Unfortunately, it also requires to solve a system of non-linear equations at each time step, which dramatically decreases the overall performance and makes it essentially unusable, at least with our implementation, for systems with a number of bins $N$ larger than a few hundred.

As for the explicit and implicit-explicit Euler methods, unfortunately, our tests showed that the logarithmic term of the ideal gas, which diverges as the density goes to zero, does not make it possible to exploit the improved stability of the latter. Common techniques such as logarithmic regularisation~\cite{li2021stability} did not help. As a result, in both schemes we have to rely on very small values of the integration time step. Moreover, the scaling of the Fourier scheme is bound by the performance of the Fast-Fourier Transform method, which is $\mathcal{O}(N \log N)$. Therefore, for the rather large number of bins we use for the 2D simulations, where $N \approx 10^6$, the simpler explicit Euler scheme, whose performance scales as $\mathcal{O}(N)$, yields a four-fold speed-up and no sensibly worse quality of the results. The code we used to run the simulations is freely available online~\cite{lorenzo_rovigatti_2024_14576360}.\\
As far as simulation times are concerned, we run simulations on NVIDIA A100 GPUs. To ensure phase separation starting from a homogeneous configuration, ternary mixtures of nanostars in a 2D box of size $L=1024$ bins (with $\Delta x=10$ nm) are simulated at least up to $T_{f}=500$ s (with $\Delta t \in [10^{-6}, 10^{-8}]$ s depending on the state point). Most simulations required 24 hours to conclude.

\section{A theoretical estimate for the surface tension}
\label{app:gamma}
\begin{figure}[!h]
    \centering
    \includegraphics[width=0.6\linewidth]{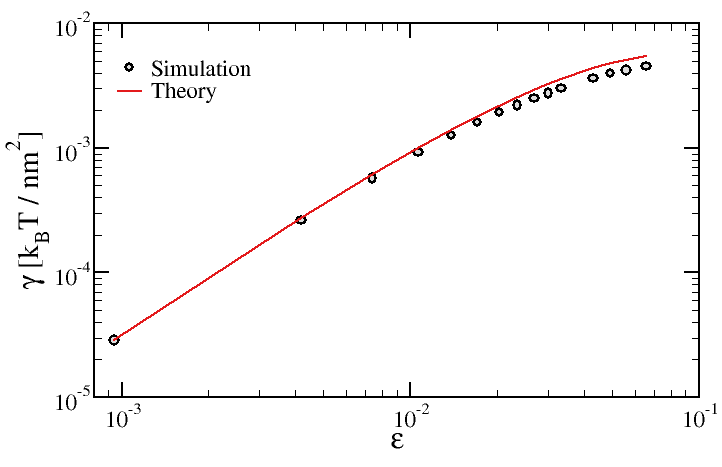}
    \caption{A direct comparison between the theoretical curve for the surface tension $\gamma(T)$ and the corresponding simulation-extracted estimates of Figure~\ref{fig:gamma_T} at $\beta K=10^6$ nm$^5$. The quantity is plotted as a function of the reduced temperature $\epsilon=\frac{T_c-T}{T_c}$, where $T_c=308.44$ K.}
    \label{fig:gamma_comparison}
\end{figure}

Our goal here is to work out a theoretical approach to compute the liquid-gas surface tension $\gamma$ as a function of temperature, and to compare this trend with the simulation-extracted data shown in Figure~\ref{fig:gamma_T}.\\
To do so, we first derive a simplified, yet general expression for the surface tension $\gamma$ starting from the 1D-free energy functional formula in Eq.~\eqref{eq:F_tot} in the one-component case:
\begin{equation}\label{0}
F(T, V) = \int_{-\infty}^{\infty} \left[ f(\rho(x)) + \frac{K}{2} \left (\frac{d \rho}{dx}\right)^2 \right] dx.
\end{equation}
The expression of the generalised chemical potential reads:
\begin{equation}
\mu=f'(\rho)-K\frac{d^2\rho}{dx^2},
\end{equation}
which can be multiplied by $\frac{d\rho}{dx}$ and set to zero (in equilibrium), yielding:
\begin{equation}
K\frac{d^2\rho}{dx^2}\frac{d\rho}{dx}=f'(\rho)\frac{d\rho}{dx}.
\label{eq:K_fprime}
\end{equation}
We integrate both sides by parts from $-\infty$ to $x$. This gives:
\begin{equation}
K\int_{\rho(-\infty)} ^{\rho(x)}  \frac{d^2 \rho'(x)}{dx^2} \frac{d\rho'}{dx}=\frac{K}{2}\left(\frac{d\rho}{dx}\right)^2,
\end{equation}
for the left-hand side term and:
\begin{equation}
\int_{\rho(-\infty)} ^{\rho(x)}  f(\rho') \frac{d\rho'}{dx} dx=f(\rho(x))-f\left(\rho\left(-\infty\right)\right),
\end{equation}
for the right-hand side term of Eq.~\eqref{eq:K_fprime}, which therefore can be written as
\begin{equation}\label{1}
\frac{K}{2}\left(\frac{d\rho}{dx}\right)^2=f(\rho(x))-f(\rho(-\infty)).
\end{equation}
A similar computation can be performed when integrating from $x$ to $\infty$, which gives:
\begin{equation}
K\int_{\rho(x)}^{\rho(\infty)}\frac{d^2 \rho'(x)}{dx^2} \frac{d\rho'}{dx}=-\frac{K}{2}\left(\frac{d\rho}{dx}\right)^2,
\end{equation}
for the left-hand side term and:
\begin{equation}
\int_{\rho(x)} ^{\rho(\infty)}  f(\rho') \frac{d\rho'}{dx} dx=f(\rho(\infty))-f(\rho(x)),
\end{equation}
for the right-hand side term, so that Eq.~\eqref{eq:K_fprime} can also be written as
\begin{equation}\label{2}
\frac{K}{2}\left(\frac{d\rho}{dx}\right)^2=f(\rho(x))-f(\rho(\infty)).
\end{equation}
Summing Eqs.~\eqref{1} and~\eqref{2}, we obtain:
\begin{equation}\label{3}
K\left(\frac{d\rho}{dx}\right)^2=2f(\rho(x))-f(\rho(\infty))-f(\rho(-\infty)).
\end{equation}
The surface tension $\gamma$ is defined as in Eq.~\eqref{eq:gamma}, \textit{i.e.}:
\begin{equation}
\gamma=\int_{-\infty}^{0}\left(f(\rho(x))-f\left(\rho\left(-\infty\right)\right)+\frac{K}{2}\left(\frac{d\rho}{dx}\right)^2\right)dx+\int_{0}^{\infty}\left(f(\rho(x))-f\left(\rho\left(\infty\right)\right)+\frac{K}{2} \left(\frac{d\rho}{dx}\right)^2\right)dx.
\end{equation}
Since $\int_{-\infty}^{\infty}f(\rho(\infty))dx=2\int_{0}^{\infty}f(\rho(\infty))dx$ and $\int_{-\infty}^{\infty}f(\rho(-\infty))dx=2\int_{-\infty}^{0}f(\rho(-\infty))dx$, it follows that:
\begin{equation}
\gamma=\int_{-\infty}^{\infty}\left(f\left(\rho\left(x\right)\right)+\frac{K}{2} \left(\frac{d\rho}{dx}\right)^2-\frac{1}{2}f(\rho(\infty))-\frac{1}{2}f(\rho(-\infty))\right)dx,
\end{equation}
and using Eq.~\eqref{3}:
\begin{equation}\label{4}
\gamma=\int_{-\infty}^{\infty}  K \left(\frac{d\rho}{dx}\right)^2dx,
\end{equation}
which is an expression of general validity (similar derivations for this equation can be found in~\cite{zwicker2025}).\\
We now make the reasonable assumption that the equilibrium density profile is described by a hyperbolic tangent function interpolating between the coexisting densities $\rho(\infty)$ and $\rho(-\infty)$ through a smooth interface with a width controlled by the parameter $\alpha$; it takes the following expression:
\begin{equation}\label{5}
\rho(x)=\frac{\rho(\infty)+\rho(-\infty)}{2} + \frac{\rho(\infty)-\rho(-\infty)}{2}\tanh{(\alpha x)}.
\end{equation}
By substituting the expression for $\rho(x)$ of Eq.~\eqref{5} in Eq.~\eqref{4}, we get an integral which is analytically solvable:
\begin{equation}\label{6}
\gamma=K\frac{\left(\rho(\infty)-\rho(-\infty) \right)^2}{4} \alpha^2 \int_{-\infty}^{\infty}\frac{1}{\cosh^4{\alpha x}}dx=K\frac{\left(\rho(\infty)-\rho(-\infty) \right)^2}{3} \alpha,
\end{equation}
where we notice that $\gamma$ depends linearly on $\alpha$.\\
We now focus on our system of interest and identify $f(\rho)=f_{\rm{wertheim}}(\rho)$, $\rho(\infty)=\rho_{\rm{liquid}}(T)$ and $\rho(-\infty)=\rho_{\rm{gas}}(T)$. We know that at equilibrium $\alpha$, which as stated above controls the width of the interface between the two coexisting densities, is equal to $\alpha_{\textit{eq}}$, for which the total free energy $F$ of the system in Eq.~\eqref{0} is minimal. From Eq.~\eqref{6}, it therefore follows that it suffices to work out (in our case numerically, since the homogeneous free energy term $f_{\rm{wertheim}}(\rho(x))$ cannot be analytically integrated) the law $\alpha_{\textit{eq}}(K, T)$ near the critical point to obtain $\gamma(K, T)$ in the same region.\\
The integral in Eq.~\eqref{0} for the total free energy has been calculated for different values of $\alpha$, thus sampling the parameter phase space, and the value of $\alpha_{\textit{eq}}(T)$ tat minimises the integral has been selected for each of the investigated values of temperature $T$ at constant $\beta K=10^6$ nm$^5$. This procedure allows to reconstruct numerically the law $\alpha_{\textit{eq}}(T)$, which can then be used in Eq.~\eqref{6} to work out $\gamma(T)$.\\
 Figure~\ref{fig:gamma_comparison} shows a direct comparison between the theoretical curve $\gamma(T)$ obtained in this way and the simulation-extracted data in Figure~\ref{fig:gamma_T} plotted as a function of the reduced temperature $\epsilon=\frac{T_c-T}{T_c}$, showing an excellent agreement between the two approaches close to the critical point. As the temperature decreases, a difference between the theoretical and numerical data arises, likely because of the relatively strong assumption of a symmetric density profile at equilibrium made in Eq.~\eqref{5}. Indeed, the latter is known to hold close to the critical point, where the free energy takes on its universal form, but not necessarily away from it.
\newpage
\bibliography{main}

\end{document}